\documentclass{osa-article}
\journal{oe}
\articletype{Research Article}
\usepackage{upgreek}
\newcommand{\enquote}[1]{``#1''}

\newcommand{\dois}[2]{\href{https://doi.org/#1}{#2}}
\newcommand{\hreff}[2]{\href{#1}{#2}}

\begin{document}

\title{Cold-atom clock based on a diffractive optic}
\author{R.\ Elvin,$^1$ G.~W.\ Hoth,$^1$ M.\ Wright,$^1$ B.\ Lewis,$^1$ J.~P.\ McGilligan,$^{2,3}$ 
A.~S.\ Arnold,$^{1*}$ P.~F.\ Griffin$^1$, and E.\  Riis$^1$}

\address{$^1$Department of Physics, SUPA, University of Strathclyde, Glasgow, G4 0NG, UK}
\address{$^2$National Institute of Standards and Technology, Boulder, Colorado 80305, USA}
\address{$^3$University of Colorado, Department of Physics, Boulder, Colorado 80309, USA}
\email{$^*$aidan.arnold@strath.ac.uk}


\ociscodes{(020.1335) Atom optics; (120.3940) Metrology; (020.3320) Laser cooling; (120.3180) Interferometry.}


\begin{abstract}
Clocks based on cold atoms offer unbeatable accuracy and long-term stability, but their use in portable quantum technologies is hampered by a large physical footprint. Here, we use the compact optical layout of a grating magneto-optical trap (gMOT) for a precise frequency reference. The gMOT collects $10^7$ $^{87}$Rb atoms, which are subsequently cooled to $20\,\upmu$K in optical molasses. We optically probe the microwave atomic ground-state splitting using lin$\perp$lin polarised coherent population trapping and a Raman-Ramsey sequence. With ballistic drop distances of only $0.5\,$mm, the measured short-term fractional frequency stability is $2 \times 10 ^{-11} /\sqrt{\tau}$.
\end{abstract}




\section{Introduction}

Cold atoms hold great promise as a medium in which ultralow-drift metrology can be achieved with excellent precision across a wide range of relevant physical observables, e.g.\ time \cite{ludlow_2015}, acceleration \cite{kasevich_2013}, rotation \cite{landragin_2016} and magnetic fields \cite{stamper_2007}. Time-keeping in particular has seen dramatic progress through cold-atom optical clocks \cite{ludlow_2015,peik,campbell_2017,murray_2018,ludlow_2018,brewer_2019,ye_2019} which now average down as fast as $\approx 10 ^{-16}/\sqrt{\tau}\;$\cite{ludlow_2018,ye_2019}, to base sensitivities of $\approx 10 ^{-18}\;$ \cite{ludlow_2018,ye_2019} in hours, enabling geodetic applications and searches for the variation of fundamental constants \cite{ludlow_2015}. 

These experimental machines are necessarily bulky due to the plethora of high-specification components required. 
However, important progress has been made towards reducing the footprint of atomic clocks: portable cold-atom optical clocks have been developed with multi-cubic-metre scale and instabilities $(1-7)\times10^{-15}/\sqrt{\tau}\;$  \cite{koeller_2017,grotti_2018};  commercial cold-atom microwave clocks still have volumes at least $4\times10^4\,$cm$^3\;$ \cite{cRb} and average down at $(3-8)\times 10^{-13}/\sqrt{\tau}\;$ \cite{muquans,cRb}; whereas `warm' chip scale atomic clocks (CSACs \cite{knappe_2004})  reach $<17\,$cm$^3$ volume and sensitivities $\approx 10 ^{-10} /\sqrt{\tau}$ but, importantly, have a relative accuracy aging rate of $\approx 10^{-9}$/month$\;$ \cite{microsemi,microsemi2}.

An atomic clock benefiting from laser cooling -- with drastically reduced size, weight, power and complexity -- would prove valuable for operations outside the laboratory environment \cite{hub} and act as an important stepping stone toward the realisation of many precision sensors. 

\begin{figure}[!t]
	\centering
	\includegraphics[width=.6\columnwidth]{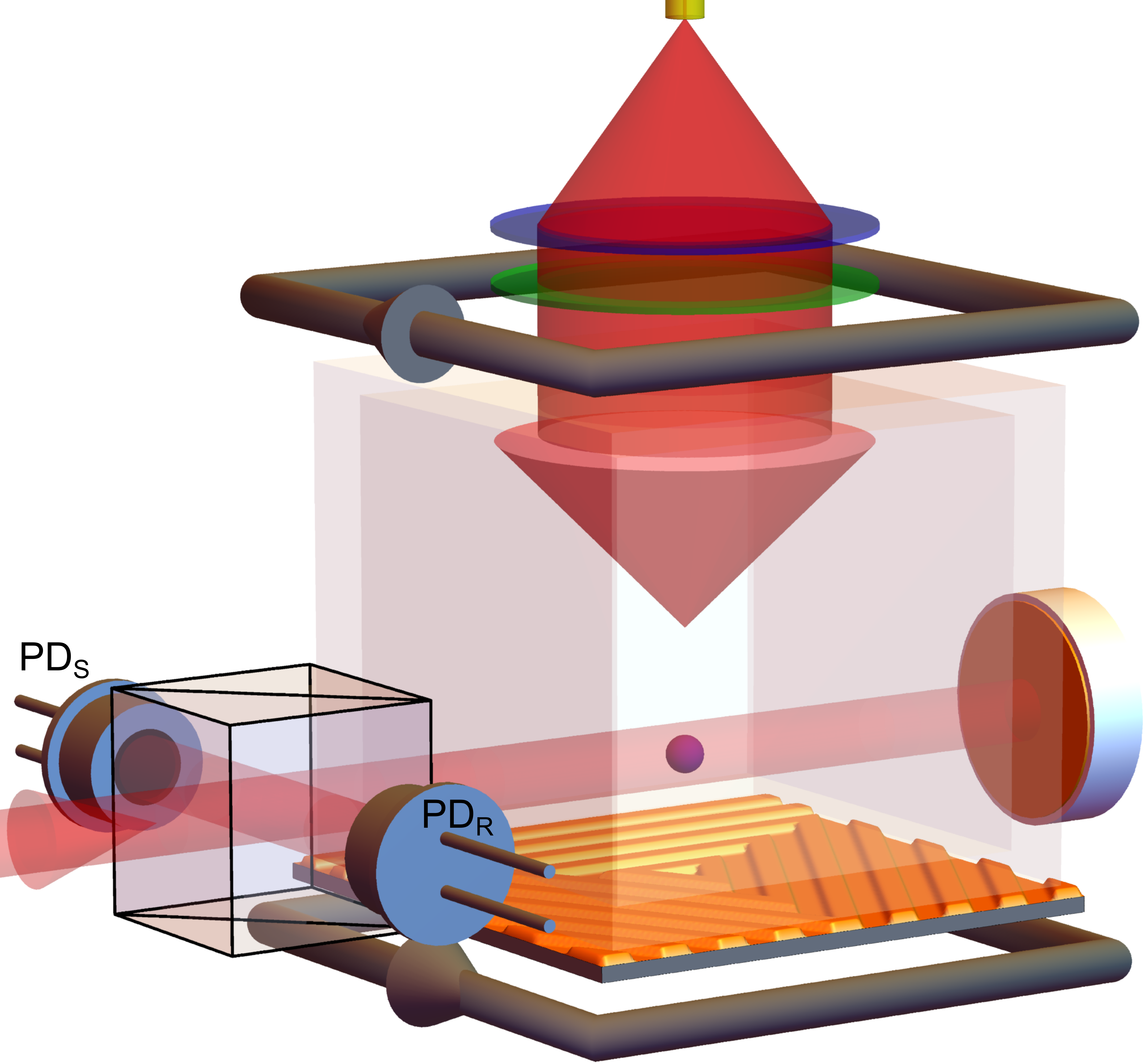}
	\caption{Schematic of the gMOT with CPT detection. 
	The single beam required for the gMOT (red downward arrow) is produced by collimating the output of an optical fiber with a lens (blue disk). It is circularly polarized by a quarter-wave plate (green disk), and directed to a $2\times 2\,$cm$^2$ microfabricated diffractive optic (gold grating), which creates the remaining three beams. All four beams intersect to make up the trapping volume within a vacuum chamber represented by the transparent cuboid. A set of anti-Helmholtz coils (grey) co-axial with the incident trapping beam generate the required 3D magnetic field gradient.
	The clock signal is obtained by measuring the transmission of the probe D$_1$ light (red arrow, left) through the cold atoms (purple ball) using the two photodiodes (PD$_\textrm{R/S}$ for reference/signal). The beamsplitter cube splits 90/10 (reflecting/transmitting -- a convention we use throughout this paper) and is non-polarising.}
	\label{fig_schematic}
\end{figure}

Grating magneto-optical traps (gMOTs) show great potential as a cold-atom source for integrated, portable devices \cite{nshii_2013}. From a single input trapping laser beam (Fig.\ \ref{fig_schematic}), gMOTs have demonstrated trapped atom numbers $\approx 10^8$ comparable to regular 6-beam MOTs with the same beam overlap volume. They can also reach sub-Doppler optical molasses temperatures, with $3\,\upmu$K measured in $3\times 10^6$ Rb atoms \cite{mcgilligan_2017} -- comparable to the best achieved in 6-beam Rb \cite{gibble_1998} and pyramidal  Rb/Cs \cite{lee_1996,bodart_2010,mueller_2017}  MOTs. Moreover, the gratings can be utilised both inside and outside the vacuum, and the gMOT is a single-input-beam MOT geometry suited for both large atom number $\gg 10^9$ and low temperature \cite{vangeleyn_2009,vangeleyn_2010}. 

Here we present an $^{87}$Rb gMOT-based cold-atom microwave clock \cite{sesko_wieman_1989} (Fig.\ \ref{fig_schematic}), as a test-bed to investigate the combination of the gMOT with coherent population trapping (CPT) using pulsed Ramsey-like probing. With the removal of the additional five cooling beams constraining the experiment volume, and a flat diffractive optic, the gMOT can be miniaturised without significantly compromising optical access.

We use the CPT technique to apply the microwave interrogation of our atoms optically \cite{vanier_CPTreview_2005}, to support our goal of a compact physics package with good optical access  without the constraints imposed by a microwave cavity.
In the CPT approach, a single beam with two phase-coherent components is used to drive the atoms into a coherent superposition of the hyperfine ground states. When the relative detuning of the two frequency components matches the hyperfine ground-state splitting, the atoms become decoupled from the light field. This leads to a sharp peak in the laser transmission, which can be used as a frequency reference, and the technique was important in the development of compact atomic clocks such as CSAC \cite{knappe_2004}. There are now several high-contrast CPT polarisation schemes including: lin$\parallel$lin \cite{esnault_2013,liu_lowdrift_2017}, lin$\perp$lin \cite{zanon_2005},  
push-pull optical pumping (PPOP) \cite{abdel_2017},  
and $\sigma_{+} - \sigma_{-}$ \cite{Kargapoltsev,liu_yudin_taichenachev_kitching_donley_2017,elgin_2019}. 
These techniques have been applied to both thermal vapour-cell and laser-cooled atoms. Vapour-cell clocks which utilise high-contrast CPT have achieved excellent short-term stability, reaching a few parts in $10 ^{-13}$ in one second  \cite{danet_dickeffect_2014,abdel_2017}. 
 However, long-term stability is challenging due to temperature sensitivity, and the buffer gases and cell-wall coatings used to increase atomic coherence time. 

With cold-atom CPT, it has been difficult to reach the same level of short-term stability as the vapour-cell CPT systems, but cold systems offer the potential for good long-term averaging \cite{liu_yudin_taichenachev_kitching_donley_2017} and improved immunity to drifts -- as seen in e.g.\ cold-atom fountain clocks \cite{peil_2017}.
Furthermore, cold atoms are a clean test bed for measuring the fundamental frequency shifts associated with CPT interrogation  \cite{blanshan_lightshifts_2015,liu_yudin_taichenachev_kitching_donley_2017,pollock_acstark_2018}.
The highest reported fractional frequency stability in a cold-atom CPT experiment to date, $1.3 \times 10 ^{-11} /\sqrt{\tau}$ averaging down to $2 \times 10 ^{-13}$ after 11 hours   \cite{liu_yudin_taichenachev_kitching_donley_2017}, utilises a $\sigma_{+} - \sigma_{-}$ polarisation configuration. In our experiment, also with ultra-cold $^{87}$Rb  but employing single-laser lin$\perp$lin CPT, we achieve comparable short-term stability of $2 \times 10 ^{-11} /\sqrt{\tau}$, with a compact gMOT rather than the conventional 6-beam MOT.

\section{Experimental methods}

The gMOT we use for the clock is similar to our previous work \cite{nshii_2013,mcgilligan_phasespace_2015,mcgilligan_2017}. We prepare 10$^{7}$ atoms in the gMOT using trap and repump light from two external cavity diode lasers (ECDLs) \cite{ecdl_1998},  
stabilised with saturated absorption spectroscopy (SAS). We achieve a large atom number with a short load time of $80\,$ms by recapturing the cold atoms between experimental cycles  \cite{sandia,riis_SPIE_2019}. The $2\times 2\,$cm$^2$  microfabricated grating chip  \cite{knt}, consists of three linear gratings with the grooves forming equilateral triangles, and is situated beneath the glass vacuum chamber (Fig.\ \ref{fig_schematic}). 

A single beam of cooling and repumping light is collimated, circularly polarised and directed onto the grating chip. The inward diffracted orders and incident beam overlap to create the atom trapping volume \cite{nshii_2013,mcgilligan_phasespace_2015}. Two anti-Helmholtz coils co-axial with the input beam (Fig. \ref{fig_schematic}) are used to generate a quadrupole magnetic field with $15\,$G/cm axial gradient that positions the MOT $\approx\!5\,$mm 
above the grating centre. Helmholtz coil pairs arranged around the chamber zero the ambient magnetic field at the location of the atoms. An additional set of coils are used to apply a homogeneous field along the propagation axis of the CPT probe beam, to lift the magnetic sub-level degeneracy. Our current system does not have magnetic shielding and, because we do not switch off the CPT quantisation field during molasses, this presently limits our temperature to $20\,\upmu$K \cite{mcgilligan_2017}. 

\begin{figure*}[!b]
	\centering
	\includegraphics[width=\textwidth]{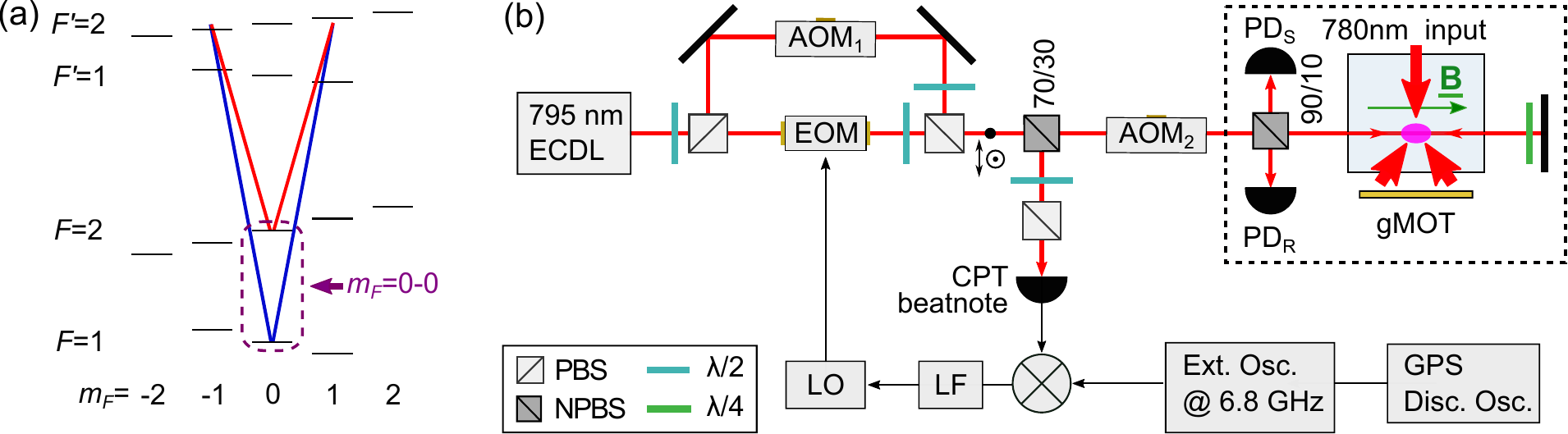}
	\caption{(a) The level diagram for lin$\perp$lin CPT on the D$_{1}$ line of $^{87}$Rb. The two orthogonally polarised CPT fields are represented by the red and blue solid lines, and the levels used in the CPT state are labelled $m_{F}=0-0$. (b) Schematic optical bench for the CPT experiment including: external cavity diode laser (ECDL);  acousto-optical modulator (AOM);  non-polarising beam-splitter (NPBS);  photodiode for reference/signal (PD$_\textrm{R/S}$); local oscillator (LO); loop filter (LF). The dashed region indicates the gMOT and CPT signal zone shown in Fig.~\ref{fig_schematic}.}
	\label{fig_expdiagram}
\end{figure*}

The relevant energy level diagram and optical layout for the CPT section of the experiment are shown in Fig.~\ref{fig_expdiagram}. A $795\,$nm ECDL is locked close to the $F=2 \rightarrow F'=2$ transition on the  D$_{1}$ line of $^{87}$Rb using SAS.  An additional AOM (not shown in figure) is used to offset the laser lock, which allows us to control the one-photon detuning. We use the $F=1\,\&\,2\rightarrow F'=2$ transitions in order to minimize  excitation of the weakly magnetic $\Delta m=2$ CPT resonances which are formed between $|F=1,m_F=\pm1\rangle$ and $|F=2,m_F=\mp1\rangle$ \cite{elvin_EFTF_2018,warren_experimental_2017,boudot_2009}. This concentrates the atomic population in the $|F=1\,\&\,2,m_F=0\rangle$ ground states and minimises the magnetic sensitivity of the clock signal amplitude.

The fields originate from the same laser, and as such they are intrinsically phase coherent. However, splitting and re-combining the optical path increases the sensitivity to small variations in the path lengths \cite{li_2017}. To counter this, we implement a phase-locked loop (PLL) in order to suppress the additional phase noise between the CPT field components. We compare the phase of the CPT components in our probe by measuring a beatnote between light picked off by the 70/30 NPBS in Fig.~\ref{fig_expdiagram}(b) and a commercial microwave source operating at a frequency close to the ground-state splitting frequency. This is used as the master oscillator to phase-lock our local oscillator (LO) and reduce the additional phase noise. Further information on the PLL can be found in \cite{riis_SPIE_2019}. 

The LO sets the EOM frequency, and the EOM RF power is set such that each first-order sideband contains $\approx 30\%$ of the power in the modulated beam. A final AOM (AOM$_2$ in Fig.~\ref{fig_expdiagram}(b) is then used for fast switching of the beam intensity for the Ramsey sequence, and to bring both beams close to resonance with the one-photon detuning set by the AOM in the SAS. The final frequency shift brings the relevant CPT fields into two-photon resonance with the $|F= 1,2\rangle \rightarrow |F' = 2\rangle$ hyperfine transitions shown in Fig.~\ref{fig_expdiagram}(a). Additional off-resonant frequency components are also present, corresponding mainly to the carrier and the other sideband from the EOM path, and we therefore expect alternative frequency preparation schemes would halve the background signal of the clock. Frequency sources related to CPT in the experiment are referenced to a GPS disciplined oscillator (\hreff{https://endruntechnologies.com/products/time-frequency/gps-frequency-standard}{Meridian II}), 
which has a specified instability of less than $1 \times 10 ^{-12}$.

The CPT probe light after AOM$_2$ is then used to interrogate the atoms. The beam is retro-reflected through the cloud to minimise Doppler shifts from the atom-light interaction \cite{esnault_2013,doppler}. 
The single-pass optical intensity is approximately $50\,\upmu$W/cm$^2$ in each resonant CPT component. 
Each pass 
through the cloud drives $m_F=0-0$ CPT transitions, and the atomic coherence forms a standing wave with a spatial period -- as a function of the  retro-reflecting mirror position -- of half the $44\,$mm microwave hyperfine splitting wavelength \cite{esnault_2013,yudin}. A translation stage is used for fine control of the reflection mirror to maximise the Ramsey fringe amplitude  at the position of the cold-atom cloud.  

One $100\,$ms 
experimental sequence consists of a stage to load the gMOT and cool the atoms in optical molasses, followed by the Ramsey-like two-pulse probing sequence \cite{zanon_2005,thomas_1982}. The first pulse pumps atoms into the dark state and is set to be 300$\,\upmu$s long in order to ensure the pumping reaches equilibrium. We then let the atoms freely precess for a variable time $T$ to build up a phase between the atomic resonance frequency and the CPT fields. A second detection pulse is used to map the accrued phase into absorption and detect the atoms' state. 
We characterize the CPT resonances in our system by measuring the steady-state transmission of the first pulse as a function of the LO frequency. We measure Ramsey-CPT fringes by integrating the initial transient in the second detection pulse, with laser intensity noise suppressed using a double-ratio method \cite{esnault_2013, liu_lowdrift_2017,liu_yudin_taichenachev_kitching_donley_2017,riis_SPIE_2019}.

The first stage of the double-ratio uses a reference photodiode (PD$_\textrm{R}$ in Fig.~\ref{fig_expdiagram}(b) to detect one output of the 90/10 NPBS before the rest of the light is sent through the atoms and detected on the signal photodiode PD$_\textrm{S}$. We take the ratio of the two signals,  PD$_\textrm{S}$/PD$_\textrm{R}$, to cancel intensity noise common to both photodiodes.

The second stage of noise cancellation exploits the time-dependent structure of the Ramsey-CPT signal. In the detection pulse, information about the atomic state resides in the transient at the beginning of the pulse. 
By taking the ratio of the integrated transmission of the signal region at the beginning of the second pulse with the integrated transmission at the end of the second pulse, we improve the intensity noise cancellation. We find an optimal integration region of 26$\,\upmu$s at the beginning of the second pulse to maximise the signal:noise ratio (SNR) of the detected fringes.

\section{CPT fringes and Allan deviation}

\begin{figure}[!t]
	\centering
	\includegraphics[width=.5\columnwidth]{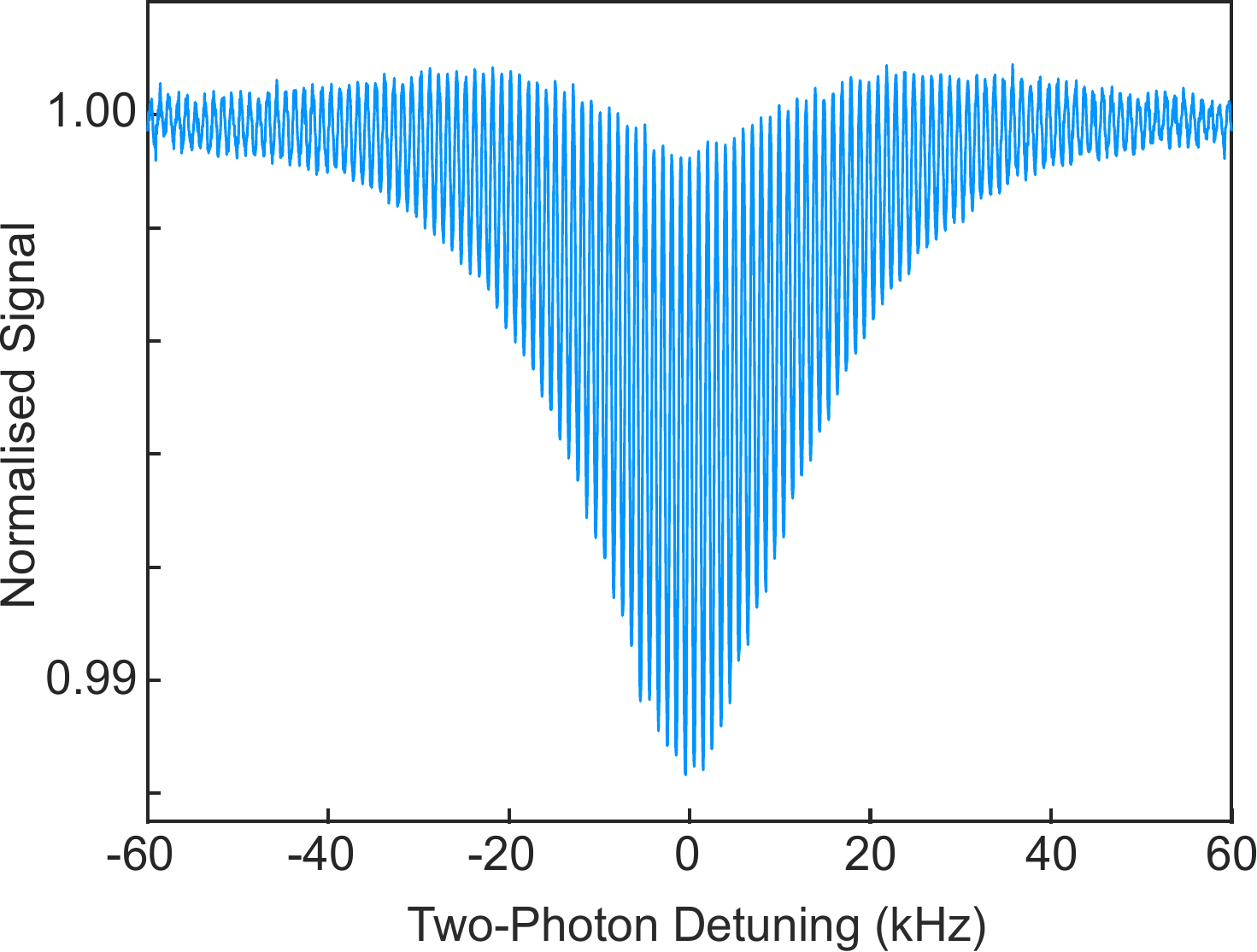}
	\caption{Experimental Ramsey-CPT fringes with a $T=1\,$ms free evolution time. We use the double-ratio technique, described in the text, to cancel common-mode laser intensity noise. Each data point represents a single run of the experiment sequence as the LO frequency is varied. The signal is normalised to the off-resonant wings. 
	}
	\label{fig_fringe}
\end{figure}

With the retro-reflected configuration, the maximum one-photon absorption is around 5\% with relative CPT contrast of 65-70\%. 
An example of the Ramsey fringes obtained using the double-ratio technique is shown in Fig.~\ref{fig_fringe}. 
These fringes were observed with a free evolution time $T=1\,$ms, corresponding to a fringe width of $1\,$kHz. Each data point represents a single run of the experiment sequence, with an approximate SNR of 100. Without the second ratio we observe the fringes as a modulation of the CPT transmission peak, with decreasing fringe contrast as the two-photon detuning moves away from resonance. The outcome from the second normalisation is observed in the shape of the Ramsey-CPT envelope in Fig.~\ref{fig_fringe} 
 \cite{liu_lowdrift_2017}.

We characterise the frequency stability of our apparatus using the CPT-Ramsey fringes. Alternating between the sides of the central fringe is a common method of obtaining the magnitude and direction of variations between the LO and the atomic resonance frequency. 
To measure the short-term stability of the apparatus an Allan deviation is calculated after approximately an hour of measurement  time. The free evolution time is set to $T=10\,$ms, which corresponds to a fringe width of $100\,$Hz.

\begin{figure}	
	\centering
	\includegraphics[width=.5\columnwidth]{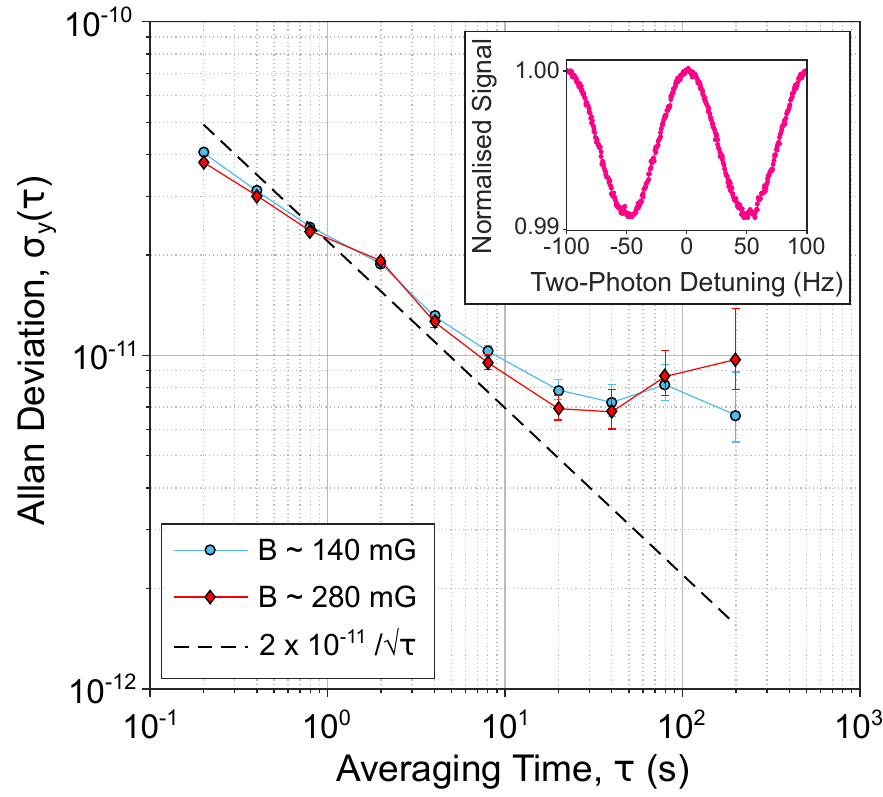}
	\caption{Measured Allan deviation curves for our apparatus. Experimental data, with a $T=10\,$ms free evolution time and PLL engaged, are plotted for two CPT bias field configurations: $280\,$mG (red diamonds) and $140\,$mG (blue circles). The dashed line represents $2 \times 10 ^{-11}/ \sqrt{\tau}$, which is a fit to the first six points. Inset: an example $T=10\,$ms fringe, where each point is a single experimental run.}
	\label{fig_fringe_ADEV}
\end{figure}

Fig.~\ref{fig_fringe_ADEV} shows standard Allan deviation plots with our clock apparatus at two different bias field magnitudes, $B\approx140\,$mG  and $280\,$mG, with an example $T=10\,$ms fringe plotted in the inset. There is little difference between the curves at the two bias fields, indicating that magnetic field instabilities are not currently a limiting factor. In the development of the apparatus, we found that triggering the experiment on the mains AC-line reduced $50\,$Hz noise contributions. 
We have measured the short-term stability to be $2 \times 10^{-11}/ \sqrt{\tau}$ (dashed line in Fig~.\ref{fig_fringe_ADEV}). 
Further experiments have shown that the degradation in stability around $100\,$s coincides with the air conditioning cycle and corresponding changes in the lab temperature.

The measured signal-to-noise ratio of $60-80$ on the side of our CPT Ramsey fringes was within a factor of two of the detection noise floor. This floor is limited by a combination of residual laser intensity, detection electronic and photon shot noise. These technical obstacles will have to be overcome in order to improve the short-term stability. In terms of the experiment's fundamental limits, using our evolution time of $T=10\,$ms, and $10^7$ atoms per $100\,$ms cycle, the quantum projection noise limit is $2\times 10^{-13}/\sqrt{\tau}\;$ \cite{Santarelli}.

To fully develop the clock system, we will implement $\upmu$-metal shielding around the chamber to both control magnetic bias fields and eliminate gradients near the atoms. We will also investigate other ways of measuring the CPT signal, such as fluorescence detection \cite{xi_2010}. In recent years there have been several developments of different pulsed schemes such as generalised auto-balanced Ramsey, which apply interleaved Ramsey sequences for suppression of probe-induced frequency shifts \cite{yudin_combined_2018, yudin_generalized_2018}, which would be beneficial to implement in our system. 

\section{Conclusion}

We have presented an $^{87}$Rb CPT clock based on grating chip laser-cooling -- and therefore compact optical design. Cold atom quantum technology has the potential for long-term accuracy without the drift often seen in compact thermal systems. 
With lin$\perp$lin CPT in a Raman-Ramsey scheme we measured a short-term frequency stability of $2\times 10 ^{-11}/ \sqrt{\tau}$, 
with upgrades planned in future to tackle frequency stability limitations in the mid- to long-term, and prospects for $2 \times 10 ^{-13} /\sqrt{\tau}$ at the quantum projection noise limit. Our optically compact cold-atom frequency reference is a valuable stepping stone toward the realisation of portable precision metrological devices with significantly reduced size, weight and power. The dataset used in this paper is available here \cite{dataset}.

\section*{Funding}

EPSRC (EP/M013294/1); the National Physical Laboratory; DSTL (DSTLX-100095636R).

\section*{Acknowledgments}

We greatly appreciated discussions with Leo Hollberg, Elizabeth Donley, Gaetano Mileti, Christoph Affolderbach, and Florian Gruet.


\begin{thebibliography}{10}

\bibitem{ludlow_2015}
A.~D. Ludlow, M.~M. Boyd, J.~Ye, E.~Peik, and P.~O. Schmidt, \enquote{Optical atomic clocks,} \dois{10.1103/RevModPhys.87.637}{Rev.\ Mod.\ Phys.\ \textbf{87}, 637--701 (2015)}.

\bibitem{kasevich_2013}
S.~M.\ Dickerson, J.~M.\ Hogan, A.\ Sugarbaker, D.~M.~S.\ Johnson, and M.~A.\ Kasevich, 
\enquote{Multiaxis inertial sensing with long-time point source atom interferometry,} 
\dois{10.1103/PhysRevLett.111.083001}{Phys.\ Rev.\ Lett.\ \textbf{111}, 083001 (2013)}.

\bibitem{landragin_2016}
I.\ Dutta, D.\ Savoie, B.\ Fang, B.\ Venon, C.~L.\ Garrido Alzar, R.\  Geiger, and A.\ Landragin, 
\enquote{Continuous cold-atom inertial sensor with 1nrad/sec rotation stability,} 
\dois{10.1103/PhysRevLett.116.183003}{Phys.\ Rev.\ Lett.\ \textbf{116}, 183003 (2016)}.

\bibitem{stamper_2007}
M.\ Vengalattore, J.~M.\ Higbie, S.~R.\ Leslie, J.\ Guzman, L.~E.\ Sadler, and D.~M.\ Stamper-Kurn, 
\enquote{High-resolution magnetometry with a spinor Bose-Einstein condensate,}  
\dois{10.1103/PhysRevLett.98.200801}{Phys.\ Rev.\ Lett.\ \textbf{98}, 200801 (2007)}.

\bibitem{peik}   
N.\ Huntemann, C.\ Sanner, B.\ Lipphardt, Chr.\ Tamm, and E.\ Peik, \enquote{Single-ion atomic clock with $3\times10^{-18}$ systematic uncertainty,} \dois{10.1103/PhysRevLett.116.063001}{Phys.\ Rev.\ Lett.\ \textbf{116}, 063001 (2016)}.

\bibitem{campbell_2017}  
S.\ L.\ Campbell, R.\ B.\ Hutson, G.\ E.\ Marti, A.\ Goban, N.\ Darkwah Oppong, R.\ L.\ McNally, L.\ Sonderhouse, J.\ M.\ Robinson, W.\ Zhang, B.\ J.\ Bloom, and J.\ Ye, \enquote{A Fermi-degenerate three-dimensional optical lattice clock,} \dois{10.1126/science.aam5538}{Science \textbf{358}, 90--94 (2017)}.

\bibitem{murray_2018}
K.\ J.\ Arnold, R.\ Kaewuam, A.\ Roy, T.\ R.\ Tan, and M.\ D.\ Barrett 
\enquote{Blackbody radiation shift assessment for a lutetium ion clock,}
\dois{10.1038/s41467-018-04079-x}{Nat.\ Commun.\ \textbf{9}, 1650 (2018).}

\bibitem{brewer_2019}
S.\ M.\ Brewer, J.-S.\ Chen, A.\ M.\ Hankin, E.\ R.\ Clements, C.\ W.\ Chou, D.\ J.\ Wineland, D.\ B.\ Hume, and D.\ R.\ Leibrandt, \enquote{$^{27}$Al$^+$ quantum-logic clock with a systematic uncertainty below $10^{-18}$,}   \dois{10.1103/PhysRevLett.123.033201}{Phys.\ Rev.\ Lett.\ \textbf{123}, 033201 (2019)}.

\bibitem{ludlow_2018}
W.\ F.\ McGrew, X.\ Zhang, R.\ J.\ Fasano, S.\ A.\ Sch\"{a}ffer, K.\ Beloy, D.\ Nicolodi, R.\ C.\ Brown, N.\ Hinkley, G.\ Milani, M.\ Schioppo, T.\ H.\ Yoon, and A.\ D.\ Ludlow,
\enquote{Atomic clock performance enabling geodesy below the centimetre level,}
\dois{10.1038/s41586-018-0738-2}{Nature \textbf{564}, 87--90 (2018)}.

\bibitem{ye_2019}
E.\ Oelker, R.~B.\ Hutson, C.~J.\ Kennedy, L.\ Sonderhouse, T.\ Bothwell, A.\ Goban, D.\ Kedar,
C.\ Sanner, J.~M.\ Robinson, G.~E.\ Marti, D.~G.\ Matei, T.\ Legero, M.\  Giunta, R.\ Holzwarth, 
F.\ Riehle, U.\ Sterr, and J.\ Ye,
\enquote{Demonstration of $4.8 \times 10^{-17}$ stability at $1\,$s for two independent optical clocks,} 
\dois{10.1038/s41566-019-0493-4}{Nat.\ Photon.\  \textbf{13}, 714--719 (2019).}

\bibitem{koeller_2017}
S.~B.\ Koller, J.\ Grotti, St.\ Vogt, A.\ Al-Masoudi, S.\ D\"{o}rscher, S.\ H\"{a}fner, U.\ Sterr, and Ch.\ Lisdat, 
\enquote{Transportable optical lattice clock with $7\times10^{-17}$ uncertainty,} 
\dois{10.1103/PhysRevLett.118.073601}{Phys.\ Rev.\ Lett.\ \textbf{118}, 073601 (2017)}.

\bibitem{grotti_2018}
J.\ Grotti, S.\ Koller, S.\ Vogt, S.\ H\"{a}fner, U.\ Sterr, C.\ Lisdat, H.\  Denker, C.\  Voigt, L.\ Timmen, A.\  Rolland, F.~N.\ Baynes, H.~S.\ Margolis, M.\ Zampaolo, P.\ Thoumany, M.\ Pizzocaro, B.\ Rauf, F.\ Bregolin, A.\ Tampellini, P.\ Barbieri, M.\ Zucco, G.~A.\ Costanzo, C.\ Clivati, F.\ Levi, and D.\ Calonico, 
\enquote{Geodesy and metrology with a transportable optical clock,} 
\dois{10.1038/s41567-017-0042-3}{Nat.\ Phys.\  \textbf{14}, 437--441 (2018)}.

\bibitem{cRb}
SpectraDynamics \hreff{http://spectradynamics.com/products/crb-clock/}{cRb-Clock}.

\bibitem{muquans}
MuQuans \hreff{https://www.muquans.com/product/muclock/}{MuClock}.

\bibitem{knappe_2004}
S.~Knappe, V.~Shah, P.~D.~D. Schwindt, L.~Hollberg, J.~Kitching, L.-A. Liew, and J.~Moreland, 
\enquote{A microfabricated atomic clock,} \dois{10.1063/1.1787942}{Appl.\ Phys.\ Lett.\ \textbf{85}, 1460--1462 (2004)}.
  
\bibitem{microsemi}
CSAC specifications at \hreff{https://www.microsemi.com/product-directory/clocks-frequency-references/3824-chip-scale-atomic-clock-csac}{www.microsemi.com} (2019).

\bibitem{microsemi2}
P.\ Cash, W.\ Krzewick, P.\ Machado, K.~R.\ Overstreet, M.\ Silveira, M.\ Stanczyk, D.\ Taylor, and X.\ Zhang, 
\enquote{Microsemi chip scale atomic clock (CSAC) technical status, applications, and future plans,} 
\dois{10.1109/EFTF.2018.8408999}{IEEE Proc.\ EFTF  (2018)}.

\bibitem{hub}
K.\ Bongs, V.\ Boyer, M.~A.\ Cruise, A.\  Freise, M.\ Holynski, J.\ Hughes, A.\  Kaushik, Y.-H.\ Lien, A.\ Niggebaum, M.\  Perea-Ortiz, P.\ Petrov, S.\ Plant, Y.\  Singh, A.\ Stabrawa, D.~J.\ Paul, M.\  Sorel, D.~R.~S.\ Cumming, J.~H.\ Marsh,  R.~W.\ Bowtell, M.~G.\ Bason, R.~P.\  Beardsley, R.~P.\ Campion, M.~J.\  Brookes, T.\ Fernholz, T.~M.\ Fromhold,  L.\ Hackermuller, P.\ Kr\"{u}ger, X.\  Li, J.~O.\ Maclean, C.~J.\ Mellor, S.~V.\ Novikov, F.~Orucevic, A.~W.\ Rushforth, N.\ Welch, T.~M.\ Benson, R.~D.\ Wildman, T.\ Freegarde, M.\  Himsworth, J.\ Ruostekoski, P.\ Smith,  A.\ Tropper, P.~F.\ Griffin, A.~S.\  Arnold, E.\ Riis, J.~E.\ Hastie, D.\ Paboeuf, D.~C.\ Parrotta, B.~M.\  Garraway, A.\ Pasquazi, M.\ Peccianti,  W.\ Hensinger, E.\ Potter, A.~H.\  Nizamani, H.\ Bostock, A.\ Rodriguez Blanco, G.\ Sinuco-Leon, I.~R.\ Hill,  R.~A.\ Williams, P.\ Gill, N.\ Hempler,  G.~P.~A.\ Malcolm, T.\ Cross, B.~O.\  Kock, S.\ Maddox, and P.\ John, 
\enquote{The UK National Quantum Technologies Hub in sensors and metrology (Keynote Paper),}
\dois{10.1117/12.2232143}{SPIE Proc.\ Quantum Opt.\ \textbf{9900},   990009 (2016)}.

\bibitem{nshii_2013}
C.~C. Nshii, M.~Vangeleyn, J.~P. Cotter, P.~F. Griffin, E.~A. Hinds, C.~N. Ironside, P.~See, A.~G. Sinclair, E.~Riis, and A.~S. Arnold, \enquote{A surface-patterned chip as a strong source of ultracold atoms for quantum technologies,} \dois{10.1038/NNANO.2013.47}{Nat.\  Nanotechnol.\ \textbf{8}, 321--324 (2013)}.

\bibitem{mcgilligan_2017}
J.~P. McGilligan, P.~F. Griffin, R.~Elvin, S.~J. Ingleby, E.~Riis, and A.~S. Arnold, 
\enquote{Grating chips for quantum technologies,}
  \dois{10.1038/s41598-017-00254-0}{Sci.\ Rep.\ \textbf{7}, 384  (2017)}.

\bibitem{gibble_1998}
R.\ Legere, and K.\ Gibble, 
\enquote{Quantum scattering in a juggling atomic fountain,} 
\dois{10.1103/PhysRevLett.81.5780}{Phys.\ Rev.\ Lett.\ \textbf{81}, 5780--5783 (1998)}.

\bibitem{lee_1996}
K.~I.\ Lee, J.~A.\ Kim, H.~R.\ Noh, and W.\ Jhe, 
\enquote{Single-beam atom trap in a pyramidal and conical hollow mirror,} 
\dois{10.1364/OL.21.001177}{Opt.\ Lett.\ \textbf{21}, 1177--1179 (1996)}. 
  
\bibitem{bodart_2010}
Q.\ Bodart, S.\ Merlet, N.\ Malossi, F.\ Pereira Dos Santos, P.\ Bouyer, and A. Landragin, 
\enquote{A cold atom pyramidal gravimeter with a single laser beam,}
\dois{10.1063/1.3373917}{Appl.\ Phys.\ Lett.\ \textbf{96}, 134101 (2010)}.

\bibitem{mueller_2017}
X.\ Wu, F.\ Zi, J.\ Dudley, R.\ J.\ Bilotta, P. Canoza, and H.\ M\"{u}ller,  \enquote{Multiaxis atom interferometry with a single-diode
laser and a pyramidal magneto-optical trap,} 
\dois{10.1364/OPTICA.4.001545}{Optica \textbf{4}, 1545--1551 (2017)}.

\bibitem{vangeleyn_2009}
M.\ Vangeleyn, P.~F.\ Griffin, E.\ Riis, and A.~S.\ Arnold, 
\enquote{Single-laser, one beam, tetrahedral
magneto-optical trap,} 
\dois{10.1364/OE.17.013601}{Opt.\ Express \textbf{17}, 13601--13608, (2009)}.

\bibitem{vangeleyn_2010}
M.\ Vangeleyn, P.~F.\ Griffin, E.\ Riis, and A.~S.\ Arnold, 
\enquote{Laser cooling with a single laser beam
and a planar diffractor,}  
\dois{10.1364/OL.35.003453}{Opt.\ Lett.\ \textbf{35}, 3453--3455 (2010)}. 

\bibitem{sesko_wieman_1989}
D.~W. Sesko and C.~E. Wieman, \enquote{Observation of the cesium clock transition in laser-cooled atoms,} \dois{10.1364/OL.14.000269}{Opt.\ Lett.\ \textbf{14}, 269--271 (1989)}.

\bibitem{vanier_CPTreview_2005}
J.~Vanier, \enquote{Atomic clocks based on coherent population trapping: a review,} \dois{10.1007/s00340-005-1905-3}{Appl.\ Phys.\ B \textbf{81}, 421--442 (2005)}.

\bibitem{esnault_2013}
F.-X. Esnault, E.~Blanshan, E.~N. Ivanov, R.~E. Scholten, J.~Kitching, and E.~A. Donley, 
\enquote{Cold-atom double-$\Lambda$ coherent population trapping   clock,} \dois{10.1103/PhysRevA.88.042120}{Phys.\ Rev.\ A \textbf{88},  042120 (2013)}.

\bibitem{liu_lowdrift_2017}
X.~Liu, E.~Ivanov, V.~I. Yudin, J.~Kitching, and E.~A. Donley,
  \enquote{Low-drift coherent population trapping clock based on
  laser-cooled atoms and high-coherence excitation fields,}
  \dois{10.1103/PhysRevApplied.8.054001}{Phys.\ Rev.\ Appl.\ \textbf{8}, 054001 (2017)}.

\bibitem{zanon_2005}
T.~Zanon, S.~Gu{\'e}randel, E.~de~Clercq, D.~Holleville, N.~Dimarcq, and A.~Clairon, 
\enquote{High contrast Ramsey fringes with coherent-population-trapping pulses in a double lambda atomic system,}  \dois{10.1103/PhysRevLett.94.193002}{Phys.\ Rev.\ Lett.\ \textbf{94},   193002 (2005)}.

\bibitem{abdel_2017}
M.~Abdel~Hafiz, G.~Coget, P.~Yun, S.~Gu\'{e}randel, E.~de~Clercq, and R.~Boudot, 
\enquote{A high-performance Raman-Ramsey Cs vapor cell atomic   clock,} \dois{10.1063/1.4977955}{J.\ Appl.\ Phys.\ \textbf{121}, 104903 (2017)}.

\bibitem{Kargapoltsev}
S.~V.\ Kargapoltsev, J.\ Kitching, L.\  Hollberg, A.~V.\ Taichenachev, V.~L.\  Velichansky, and V.~I.\ Yudin, 
\enquote{High-contrast dark resonance in $\sigma^{+}-\sigma^-$ optical field},
 \dois{10.1002/lapl.200410107}{Laser Phys.\  Lett.\ \textbf{1}, 495 (2004)}.

\bibitem{liu_yudin_taichenachev_kitching_donley_2017}
X.~Liu, V.~I. Yudin, A.~V. Taichenachev, J.~Kitching, and E.~A. Donley,
  \enquote{High contrast dark resonances in a cold-atom clock probed with counterpropagating circularly polarized beams,}
 \dois{10.1063/1.5001179}{Appl.\ Phys.\ Lett.\ \textbf{111}, 224102 (2017)}.
 
 \bibitem{elgin_2019}
 J.~D.\ Elgin, T.~P.\ Heavner, J.\ Kitching, E.~A.\ Donley, J.\ Denney, and E.~A.\ Salim,
\enquote{A cold-atom beam clock based on coherent population trapping,} 
\dois{10.1063/1.5087119}{Appl.\  Phys.\ Lett.\ \textbf{115}, 033503 (2019)}.

\bibitem{danet_dickeffect_2014}
J.-M. Danet, M.~Lours, S.~Gu{\'e}randel, and E.~De~Clercq, \enquote{Dick effect in a pulsed atomic clock using coherent population trapping,} \dois{10.1109/TUFFC.2014.2945}{IEEE Trans.\ UFFC \textbf{61}, 567--574 (2014)}.

\bibitem{peil_2017}
S.\ Peil, T.~B.\ Swanson, J.\ Hanssen, and J.\ Taylor, 
\enquote{Microwave-clock timescale with instability on order of $10^{-17}$,} 
\dois{10.1088/1681-7575/aa65f7}{Metrologia \textbf{54}, 247--252 (2017)}.

\bibitem{blanshan_lightshifts_2015}
E.~Blanshan, S.~M. Rochester, E.~A. Donley, and J.~Kitching, \enquote{Light shifts in a pulsed cold-atom coherent population trapping clock,} \dois{10.1103/PhysRevA.91.041401}{Phys.\ Rev.\ A \textbf{91},  041401 (2015)}.

\bibitem{pollock_acstark_2018}
J.~W. Pollock, V.~I. Yudin, M.~Shuker, M.~Y. Basalaev, A.~V. Taichenachev, X.~Liu, J.~Kitching, and E.~A. Donley, \enquote{ac Stark shifts of dark resonances probed with Ramsey spectroscopy,}
 \dois{10.1103/PhysRevA.98.053424}{Phys.\ Rev.\ A \textbf{98}, 053424 (2018)}.

\bibitem{mcgilligan_phasespace_2015}
J.~P. McGilligan, P.~F. Griffin, E.~Riis, and A.~S. Arnold,
  \enquote{Phase-space properties of magneto-optical traps utilising
  micro-fabricated gratings,} \dois{10.1364/OE.23.008948}{Opt.\ Express \textbf{23}, 8948--8959 (2015)}.

\bibitem{ecdl_1998}
A.\ S.\ Arnold, J.\ S.\ Wilson, and M.\ G.\ Boshier, 
\enquote{A simple extended-cavity diode laser,} 
\dois{10.1063/1.1148756}{Rev.\ Sci.\ Instrum.\ \textbf{69}, 1236--1239 (1997)}.

\bibitem{sandia}
A.~V.\ Rakholia, H.~J.\ McGuinness, and G.~W.\  Biedermann, 
\enquote{Dual-axis high-data-rate atom interferometer via cold ensemble exchange,}
\dois{10.1103/PhysRevApplied.2.054012}{Phys.\ Rev.\ Appl.\ \textbf{2}, 054012 (2014)}.

\bibitem{riis_SPIE_2019}
G.~W. Hoth, R.~Elvin, M.~Wright, B.~Lewis, A.~S.\ Arnold, P.~F.\  Griffin, and E.~Riis, \enquote{Towards a compact atomic clock based on coherent population trapping and the grating magneto-optical trap,} 
 \dois{10.1117/12.2516612}{SPIE Proc.\ \textbf{10934}, 109342E (2019)}.

\bibitem{knt}
Commercially available at   \hreff{https://www.kntnano.com/quantum/gmotgrating/}{www.kntnano.com}.

\bibitem{warren_experimental_2017}
Z.~Warren, M.~S. Shahriar, R.~Tripathi, and G.~S.\ Pati, \enquote{Experimental and theoretical comparison of different optical excitation schemes for a compact coherent population trapping Rb vapor  clock,} \dois{10.1088/1681-7575/aa72bb}{Metrologia \textbf{54}, 418--431 (2017)}.

 \bibitem{boudot_2009}
 R.\ Boudot, S.\ Guerandel, E.\ de Clercq, N.\ Dimarcq, and A.\ Clairon, 
\enquote{Current status of a pulsed CPT Cs cell clock,} 
\dois{10.1109/TIM.2008.2009918}{IEEE Trans.\ Inst.\ Meas.\ \textbf{58}, 1217--1222 (2009)}.

\bibitem{elvin_EFTF_2018}
R.~Elvin, G.~W. Hoth, M.~W. Wright, J.~P.\ McGilligan, A.~S.\ Arnold, P.~F.\ Griffin, and E.~Riis, 
\enquote{Raman-Ramsey CPT with a grating magneto-optical trap,} 
 \dois{10.1109/EFTF.2018.8408998}{IEEE Proc.\ EFTF (2018)}.
    
\bibitem{li_2017}
W.\ Li, X.\ Pan, N.\ Song, X.\ Xu, and X.\ Lu, 
\enquote{A phase-locked laser system based on double direct modulation
technique for atom interferometry,} 
\dois{10.1007/s00340-016-6630-6}{Appl. Phys. B \textbf{123}, 54 (2017)}.

 \bibitem{doppler}
E.~A.\ Donley, F.-X.\ Esnault, E.\ Blanshan, and J.~ Kitching, 
 \enquote{Cancellation of Doppler shifts in a cold-atom CPT clock,}
 \dois{10.1109/EFTF-IFC.2013.6702061}{IEEE Proc.\ EFTF  (2013)}.

 \bibitem{yudin}
C.\ Affolderbach, S.\ Knappe, R.\ Wynands, A.~V.\ Ta\u{i}chenachev, and V.\ I.\ Yudin, 
\enquote{Electromagnetically induced transparency and absorption in a standing wave,} 
\dois{10.1103/PhysRevA.65.043810}{Phys.\ Rev.\ A \textbf{65}, 043810 (2002)}.

\bibitem{thomas_1982}
J.~E.\ Thomas, P.~R.\ Hemmer, S.\ Ezekiel, C.~C.\ Leiby, Jr., R.~H.\ Picard, and C.~R.\ Willis, 
\enquote{Observation of Ramsey fringes using a stimulated, resonance Raman transition in a sodium atomic beam,} 
\dois{10.1103/PhysRevLett.48.867}{Phys.\ Rev.\ Lett.\ \textbf{48}, 867--870 (1982)}.

\bibitem{Santarelli}
G.\ Santarelli, Ph.\ Laurent, P.\ Lemonde, A.\ Clairon, A.~G.\  Mann, S.\ Chang, A.~N.\ Luiten, and C.\ Salomon, 
\enquote{Quantum projection noise in an atomic fountain: a high stability cesium frequency standard,} 
 \dois{10.1103/PhysRevLett.82.4619}{Phys.\ Rev.\ Lett.\  \textbf{82}, 4619--4622 (1999)}.

\bibitem{xi_2010}
C.\ Xi, Y.\ Guo-Qing, W.\ Jin, and Z.\ Ming-Sheng,  \enquote{Coherent population trapping - Ramsey interference in cold atoms,} \dois{10.1088/0256-307X/27/11/113201}{Chin.\ Phys.\ Lett.\ \textbf{27}, 113201 (2010)}.


\bibitem{yudin_combined_2018}
V.~I. Yudin, A.~V. Taichenachev, M.~Y. Basalaev, T.~Zanon-Willette, T.~E. Mehlst{\"a}ubler, R.~Boudot, J.~W. Pollock, M.~Shuker, E.~A. Donley, and J.~Kitching, 
\enquote{Combined error signal in Ramsey spectroscopy of 
clock Transitions,} 
 \dois{10.1088/1367-2630/aaf47c}{New J.\ Phys.\ \textbf{20}, 123016 (2018)}.

\bibitem{yudin_generalized_2018}
V.~I. Yudin, A.~V. Taichenachev, M.~Y. Basalaev, T.~Zanon-Willette, J.~W.\ Pollock, M.~Shuker, E.~A. Donley, and J.~Kitching, \enquote{Generalized autobalanced Ramsey spectroscopy of clock  transitions,}
 \dois{10.1103/PhysRevApplied.9.054034}{Phys.\ Rev.\ Appl.\ \textbf{9}, 054034 (2018)}.
 
 \bibitem{dataset}
 Dataset DOI:  \dois{10.15129/4b93bf4c-84fe-4a0b-8e85-871884aefa6e}{10.15129/4b93bf4c-84fe-4a0b-8e85-871884aefa6e}.
 
\end{thebibliography}
\end{document}